%% file: neurips_2023.tex
\documentclass{article}

\usepackage[final]{neurips_2023}

\usepackage[utf8]{inputenc} 
\usepackage[T1]{fontenc}    
\usepackage{hyperref}       
\usepackage{url}            
\usepackage{booktabs}       
\usepackage{amsfonts}       
\usepackage{nicefrac}       
\usepackage{microtype}      
\usepackage{xcolor}         
\usepackage{graphicx}
\usepackage{amsmath}
\usepackage{amsthm}
\usepackage[ruled,vlined,linesnumbered]{algorithm2e}
\usepackage{subcaption}
\usepackage{diagbox}

\newtheorem{definition}{Definition}
\newtheorem{theorem}{Theorem}
\newtheorem{lemma}{Lemma}

\title{History Filtering in Imperfect Information Games: Algorithms and Complexity}

\author{%
  Christopher Solinas$^1$, \textbf{Douglas Rebstock}$^1$, \textbf{Nathan R. Sturtevant}$^{1,2}$, \textbf{Michael Buro}$^1$\\
  $^1$\textrm{Department of Computing Science, University of Alberta}\\
  $^2$\textrm{Alberta Machine Intelligence Institute (Amii)}\\
  \texttt{\{solinas,drebstoc,nathanst,mburo\}@ualberta.ca} \\
}

\begin{document}

\maketitle

\begin{abstract}
Historically applied exclusively to perfect information games, depth-limited search with value functions has been key to recent advances in AI for imperfect information games.
Most prominent approaches with strong theoretical guarantees require \textit{subgame decomposition}---a process in which a subgame is computed from public information and player beliefs.
However, subgame decomposition can itself require non-trivial computations, and its tractability depends on the existence of efficient algorithms for either full enumeration or generation of the histories that form the root of the subgame.
Despite this, no formal analysis of the tractability of such computations has been established in prior work, and application domains have often consisted of games, such as poker, for which enumeration is trivial on modern hardware.

Applying these ideas to more complex domains requires understanding their cost.
In this work, we introduce and analyze the computational aspects and tractability of filtering histories for subgame decomposition.
We show that constructing a single history from the root of the subgame is generally intractable, and then provide a necessary and sufficient condition for efficient enumeration.
We also introduce a novel Markov Chain Monte Carlo-based generation algorithm for trick-taking card games---a domain where enumeration is often prohibitively expensive.
Our experiments demonstrate its improved scalability in the trick-taking card game \textit{Oh Hell}.
These contributions clarify when and how depth-limited search via subgame decomposition can be an effective tool for sequential decision-making in imperfect information settings.
\end{abstract}

\input{sec/intro.tex}
\input{sec/bg.tex}
\input{sec/filter.tex}
\input{sec/filter-complexity.tex}
\input{sec/filter-ttcg.tex}
\input{sec/experiments.tex}
\input{sec/concl.tex}

\bibliographystyle{named}
\bibliography{refs}


\section{Supplementary Material}
\input{sec/extra.tex}
\input{sec/proofs.tex}


\end{document}

%% file: sec/intro.tex
\section{Introduction} \label{sec:intro}

Games are a standard model for sequential decision-making.
As the number of sequential decisions needed to play increases, the size of the game's state space can grow exponentially---quickly becoming too large to search exhaustively.
Depth-limited search navigates this issue by replacing decision points below a certain depth with a value function that captures or approximates the value of playing the subgame from that position onward.
This can massively boost scalability and has been key to several famous results in perfect information games including Checkers (\cite{schaeffer1996solving}), Chess (\cite{campbell2002deep}) and Go (\cite{silver2017mastering}).

The same idea has recently been successfully applied to certain imperfect information games such as poker (\cite{brown2019superhuman,moravvcik2017deepstack})---achieving similarly impressive results by defeating human experts.
In contrast to the perfect information setting, for which computing the value function only requires evaluating the current history, most successful depth-limited search algorithms in imperfect information games require evaluating whole sets of histories and their reach probabilities.
This information is used to generalize the concept of a subgame and its value by decomposing the game tree into \textit{public belief states}, which are analogous to belief states in Markov systems such as Hidden Markov Models (HMMs) and Partially Observable Markov Decision Processes (POMDPs).
This approach is commonly referred to as \textit{subgame decomposition}; we describe the associated history and reach probability computations as \textit{history filtering}.
Although current theory establishes the necessary information for theoretically sound depth-limited search in imperfect information games (\cite{kovarik2020value}), it is unclear how, or if, this information can be computed efficiently.
Understanding these computations provides insight into scaling search to larger, more complex imperfect information games.

In this work, we define variants of history filtering for subgame decomposition that are useful for search---called \textit{enumeration} and \textit{generation}---while also developing a suitable notion for efficiently solving them.
Efficient solutions should take at most polynomially many steps in the length of the input observation sequence.
We show that, in general, such algorithms only exist if $\texttt{P} = \texttt{NP}$.
From there, we investigate methods for efficient enumeration and generation.
First, by identifying a structural property of the game tree that is both a necessary and sufficient condition for efficient enumeration, and then by introducing a novel, unbiased generation algorithm for trick-taking card games, based on Markov Chain Monte Carlo.
Through experiments in \textit{Oh Hell}, we validate its improved scalability and highlight the potential advantages of this approach.
Our contributions advance the theory of depth-limited search in imperfect information domains.

%% file: sec/bg.tex
\section{Background} \label{sec:bg}
In this section, we summarize the concepts and algorithms related to history filtering and public belief states.

\subsection{Factored Observation Stochastic Games}
The recent introduction of Factored Observation Stochastic Games (FOSGs) (\cite{kovavrik2019rethinking}) has helped clarify fundamental concepts about decomposing public and private observations and information in multi-agent, partially-observable sequential decision problems.

An FOSG is a tuple $G = \left < \mathcal{N}, \mathcal{W}, \mathcal{P}, w^0, \mathcal{A}, \mathcal{T}, \mathcal{R}, \mathcal{O} \right >$.
$\mathcal{N} = \left \{1,...,N\right\}$ represents the set of players, $\mathcal{W}$ is the set of world states, and $\mathcal{A}$ is the set of joint actions.
$\mathcal{P} : \mathcal{W} \rightarrow 2^{\mathcal{N}}$ is the player function, which describes which players act in which world states.
$\mathcal{T} : \mathcal{W} \times \mathcal{A} \rightarrow \Delta \mathcal{W}$ is the state transition function, where $\Delta \mathcal{W}$ represents the set of probability distributions over $\mathcal{W}$.
$\mathcal{R} : \mathcal{W} \times \mathcal{A} \rightarrow \mathbb{R}^{N}$ assigns a reward to each player, and $\mathcal{O} : \mathcal{W} \times \mathcal{A} \times \mathcal{W} \rightarrow \mathbb{O}^{N+1}$ is the observation function---which maps transitions (world state-action-world state triples) to private observations for each player and a public observation common to all players.

Games start at the initial world state $w^0$.
In any world state $w \in \mathcal{W}$, player $i$ acts when $i \in \mathcal{P}(w)$. 
The joint action set $\mathcal{A} := \prod_{i \in \mathcal{N}} \mathcal{A}_i$ is defined as the product of each player's individual action sets across all $w \in \mathcal{W}$.
$\mathcal{A}_i(w) \subset \mathcal{A}_i$ denotes the legal actions for $i$ in $w$, and $\mathcal{A}(w) := \prod_{i \in \mathcal{P}(w)} \mathcal{A}_i(w)$ is the set of legal joint actions in $w$.
Play proceeds when each $i \in \mathcal{P}(w)$ chooses an action $a_i \in \mathcal{A}_i(w)$---resulting in joint action $a := (a_i)_{i \in \mathcal{P}(w)}, a \in \mathcal{A}(w)$.
The next state $w^\prime$ is sampled from $\mathcal{T}(w,a)$, while the reward is determined by evaluating $\mathcal{R}(w,a)$.
Finally, $\mathcal{O}(w,a,w^\prime)$ is factored into public and private observations as $(\mathcal{O}_{\text{priv}(1)}(w, a, w^\prime),... ,\mathcal{O}_{\text{priv}(N)}(w,a,w^\prime), \mathcal{O}_{\text{pub}}(w,a,w^\prime))$.

\subsection{Policies, Reach Probabilities, and Beliefs}
A \textbf{history} is a sequence $h := (w^0,a^0,w^1,a^1,...,w^t)$ of world states and actions for which $w^k \in \mathcal{W}, a^k \in \mathcal{A}(w^k)$, and $\mathbb{P}[\mathcal{T}(w^k, a^k) = w^{k+1}] > 0$ for $k, 0 \leq k \leq t-1$.
We refer to $|h| := t$ as the \textit{length} of $h$.
The set of all legal histories is denoted as $\mathcal{H}$.
We use the standard notation $h^\prime \sqsubseteq h$ to denote that $h^\prime$ is a \textbf{prefix history} of $h$ (i.e., $h^\prime$ is a subsequence of $h$ starting at $w^0$ and ending in a world state).
A \textbf{terminal history} $z \in \mathcal{Z}$ signifies that play has reached a world state where the game ends.
The \textbf{utility} for player $i$, $u_i : \mathcal{Z} \rightarrow \mathbb{R}$ is the sum of all rewards accumulated by $i$ over the world state and action sequence.

\begin{figure}[t]
\centering
\includegraphics{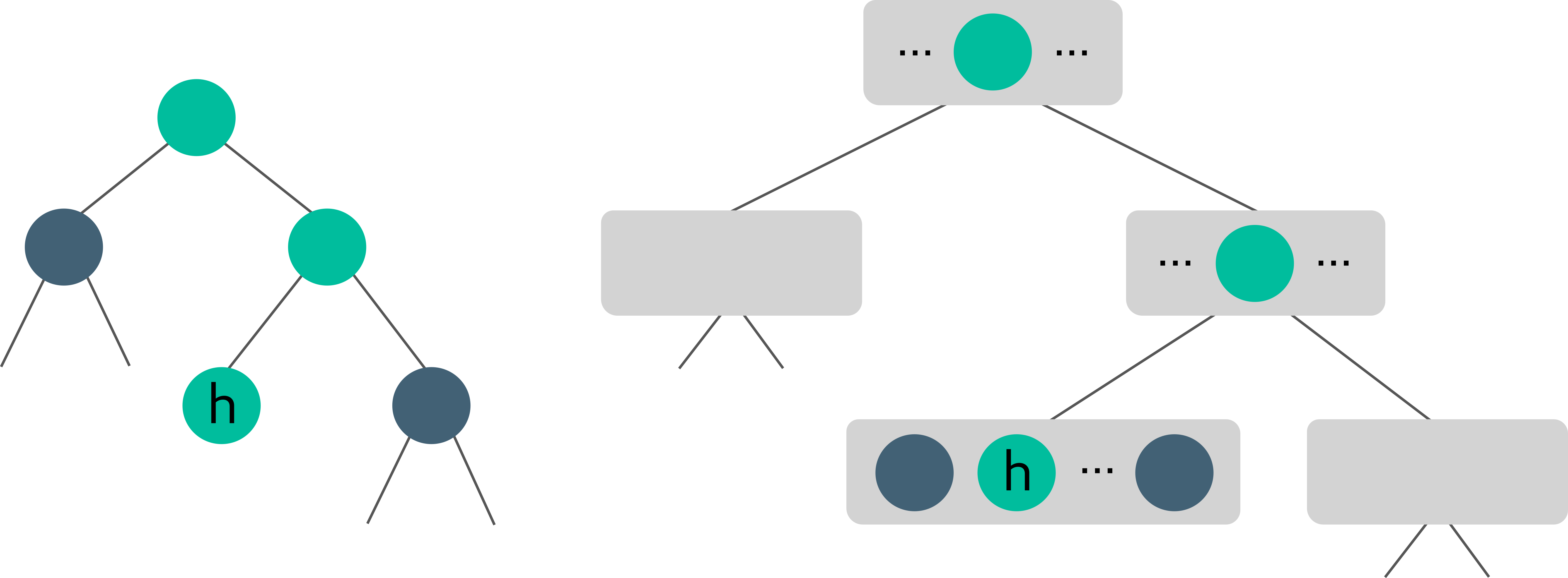} 
\caption{The same history viewed in the world state tree (left) and the public tree (right).
Grey boxes are public states, which can be consistent with many histories from the world state tree.}
\label{fig:public-tree}
\vspace{-6mm}
\end{figure}

A history $h$ produces a \textbf{public state}, which is the sequence of public observations
$s_{pub}(h) := (O^1_{pub},O^2_{pub},...,O^t_{pub})$ produced along the trajectory with $O^k_{pub} := \mathcal{O}_{pub}(w^{k-1}, a^{k-1}, w^k)$.
We denote a public state generated by an unknown history as $S \in \mathcal{S}$, where $\mathcal{S}$ is the set of all public states (and the vertex set of the \textbf{public tree}, see Figure~\ref{fig:public-tree}).
Likewise, player $i$'s \textbf{private information state} $S_i \in \mathcal{S}_i$ similarly captures the sequence of observations seen only by that player: 
$s_{i}(h) := (O^1_{i},O^2_{i},...,O^t_{i})$, where $O^k_i := \mathcal{O}_{priv(i)}(w^{k-1}, a^{k-1}, w^k)$.
Taken together, $(s_{pub}(h), s_i(h))$ represents all information available to player $i$ at history $h$, and is referred to as the player's \textbf{information state} or \textbf{infostate}.
$\mathcal{S}_i(S)$ denotes the set of all infostates consistent with public state $S$ and is a partition of $S$.
We use $\mathcal{H}_S := \{h \in \mathcal{H} : s_{pub}(h) = S\}$ to denote the set of histories consistent with public state $S$.
Since all players receive (possibly empty) observations at every world state transition, this formalism avoids \textit{non-timeability} and \textit{thick infostates} present in the extensive form.
See \citet{kovarik2020value} for details.

A player plays according to a \textbf{policy} $\pi_i : \mathcal{S}_i \rightarrow \Delta(\mathcal{A})$ which maps player $i$'s infostates to the set of probability distributions over action set $\mathcal{A}$.
A \textbf{joint policy} $\pi = (\pi_1, ..., \pi_n)$ is a tuple consisting of every player's policy.
The \textbf{reach probability} of a history under $\pi$ is $P^\pi(h) = P_c(h)P^\pi_1(h)P^\pi_2(h)...P^\pi_{N}(h)$ where each $P^\pi_i(h)$ is a product of action probabilities taken by player $i$ to reach $h$ from $w^0$, and $P_c(h)$ is the product of all probabilities from chance transitions taken according to the stochastic transition function $\mathcal{T}$.
The reach probability of an infostate (or public state) $S_i$ under $\pi$ can be expressed as $P^\pi(S_i) := \sum_{h \in S_i} P^\pi(h)$.
Infostate reach probabilities can also be decomposed into $P^\pi(S_i) = P_i^\pi(S_i)P_{-i}^\pi(S_i)$ where $-i$ denotes all players except $i$.
Since all $h \in S_i$ are indistinguishable to $i$, $P_i^\pi(S_i) = P_i^\pi(h)$, whereas $P_{-i}^\pi(S_i) = \sum_{h \in \mathcal{H}_{S_i}} P^\pi_{-i}(h)$ (\cite{kovarik2020value}).

\begin{figure}[t]
\centering
\includegraphics[width=0.7\linewidth]{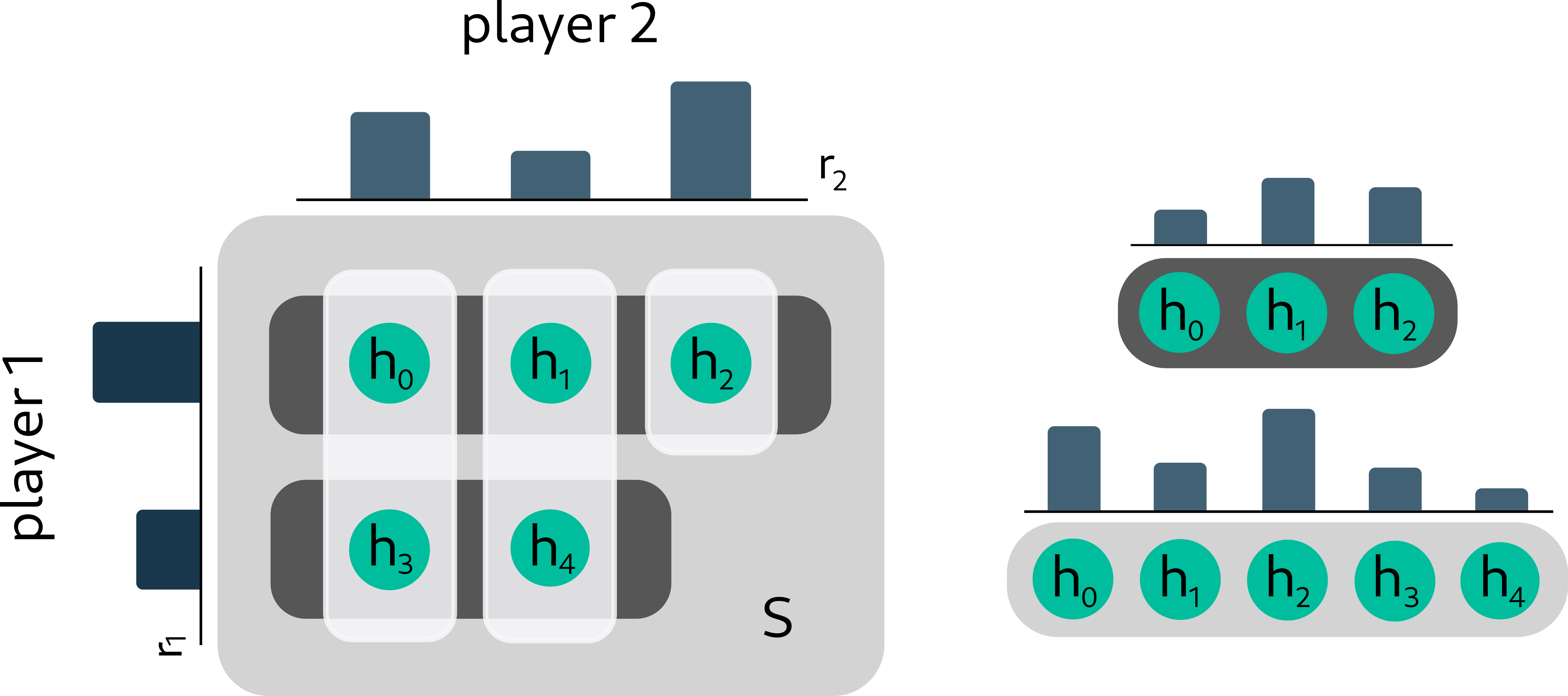}
\caption{(Left) An example public belief state in a two-player game with ranges $r_1$ and $r_2$. $S$ is consistent with a set of 5 histories. Player 1's range contains two private information states. (Right) A player's beliefs (top) and a joint range (bottom).
Player 1's beliefs, given the first private infostate, consist of reach probabilities for the individual histories in the infostate.
The joint range consists of reach probabilities over all histories in $S$.
}
\label{fig:pbs}
\end{figure}

\citet{kovarik2020value} define a \textbf{belief} under joint policy $\pi$ as the probability of reaching a history $h$ given the current infostate $S_i$: for any $h \in \mathcal{H}_{S_i}$, $P^\pi(h|S_i) = P^\pi(h) / \sum_{h^\prime \in \mathcal{H}_{S_i}} P^\pi(h^\prime)$.
The \textbf{range} for a joint policy $\pi$ at public state $S$, $r^\pi(S) := ((P^\pi_j(S_i))_{S_i \in \mathcal{S}_j(s_{pub})} )_{j=1,2,...,N}$ contains each player's reach probabilities for their infostate partition of $S$.
The (normalized) \textbf{joint range} $P^\pi(h|S)$ contains the normalized reach probabilities of all $h \in S$.
These concepts are visualized in an abstract game in Figure~\ref{fig:pbs}.
Our analysis focuses on history filtering with respect to the joint range---assuming the joint policy is constant and known by all players.

\subsection{Depth-Limited Search, Subgames, and Computing Histories}
Several depth-limited search algorithms, such as Information Set Monte Carlo Tree Search (ISMCTS) (\cite{cowling2012information}), Player of Games (\cite{schmid2021player}), and ReBel (\cite{brown2020combining}), require methods for simulating actions and state transitions. 
In these algorithms, search starts from a \textit{ground truth} history---a plausible state of the game---and proceeds until a depth limit; \textit{value functions} represent the players' expected reward in the \textit{subgame} below.

Subgames and their value functions are intuitive in the perfect information setting because they can be rooted at any non-terminal history.
However, the necessary information for computing value functions in the imperfect information setting is structured differently.
\textbf{Subgames} are instead rooted at \textbf{public belief states} (PBS), which are tuples $\beta := (S, r^\pi(S))$ composed of a public state and a range.
Searching from $\beta$ requires computing histories from $\mathcal{H}_S$ because the actions and transitions necessary for forward simulation are defined for world states, not public states, and computing optimal value functions at the subgame leaves requires ranges defined over all histories in $\mathcal{H}_S$ (\cite{kovarik2020value}).
Player of Games (\cite{schmid2021player}) opts for the Monte Carlo approach, which generates $h \in \mathcal{H}_S$ for player $i$ by sampling an infostate from $S_i \in \mathcal{S}_i(S)$ according to $i$'s component of $r^\pi(S)$, and then sampling according to their beliefs $P^\pi(h|S_i)$ (see Figure~\ref{fig:pbs} for the distinction).
ISMCTS (\cite{cowling2012information}) is similar, but instead samples directly from the normalized joint range $P^\pi(h|S)$.
Either case depends on solving a common computational problem: histories from $\mathcal{H}_S$ must be sampled according to $r^\pi(S)$, given only $S$ and $\pi$.

Belief state computation is an important topic in general Markov systems with partial observability.
As such, it appears frequently in areas such as stochastic control (\cite{nayyar2013decentralized}), learning in decentralized POMDPs (\cite{dibangoye2016optimally,oliehoek2013sufficient}), and multi-agent reinforcement learning (\cite{fickinger2021scalable}).
In FOSGs, which are a generalization of POMDPs, search techniques that use public belief states have also been crucial to the development of superhuman poker AI (\cite{brown2019superhuman,brown2020combining,moravvcik2017deepstack,schmid2021player}).
In card game AI, Perfect Information Monte Carlo (PIMC) search was successfully applied to Contract Bridge (\cite{levy1989million,ginsberg2001gib}) and later Skat (\cite{buro2009improving}).
Both applications rely on game-specific implementations for history filtering.
\citet{richards2012information} described the problem as \textit{information set generation} and provided the first generic, but exponential-time, algorithm for solving it.
\cite{seitz2021learning} provided an algorithm for approximating information set generation based on deep neural networks.
\citet{sustr2021particle} identified that exponential-sized $\mathcal{H}_S$ cause memory issues in methods that explicitly represent the range, and that efficient search requires the range to be represented compactly.

Recently, the idea of \textit{fine-tuning} has helped push the scalability of planning in POMDPs via reinforcement learning (\cite{fickinger2021scalable}) and search in FOSGs (\cite{sokota2021fine}) by eliminating the need to explicitly represent the range.
In reinforcement learning, fine-tuning consists of online updates to a parameterized model of a blueprint policy or Q-value function using sample trajectories with a fixed horizon.
These local improvements to the model helped train policies which achieved a new state-of-the-art in self-play Hanabi (\cite{fickinger2021scalable}).
Belief fine-tuning (BFT) (\cite{sokota2021fine}) extended the idea to PBS approximation without pre-training the generative model as a function of the joint policy $\pi$.

In some domains, it is unclear how to structure a dynamics model for fine-tuning such that only legal sample histories are produced.
For example, in trick-taking card games, the public observation sequence could indicate that certain players cannot possibly have certain cards in their hands; all histories output by the model should satisfy this constraint.
Our approach uses only the dynamics model defined by the game.
It is parameter-free and instead calls for simple domain-specific algorithms for history construction and neighbor generation in a Markov chain.
This results in unbiased history generation and guarantees that any sample history is legal according to the game rules.

%% file: sec/filter.tex
\section{History Computation in Public Belief States}
Our first contribution is to introduce a family of history filtering relations $\texttt{FILTER}(G, \pi)$ which formalize this computation for a given FOSG $G$ and joint policy $\pi$.
For simplicity, we limit our analysis to FOSGs with finite world state sets $\mathcal{W}$ and action sets $\mathcal{A}$, and joint policies that can be evaluated in polynomial time with respect to history length.

\subsection{Family of Filtering Relations}
Computing histories from a given public state is a special case of \textit{particle filtering}, so we name our family of relations \texttt{FILTER}. 

\begin{definition} \textbf{(FILTER)} \label{def:filter}
For any FOSG $G$ with finite $\mathcal{W}, \mathcal{A}$ and joint policy $\pi$, let $\texttt{FILTER}(G, \pi) := \{(S, h) \in \Sigma^* \times \Sigma^* : s_{pub}(h) = S, P^\pi(h|S) > 0\}$.
\end{definition}

An FOSG $G$ and a joint policy parameterize the relation---which pairs public states $S \in \mathcal{S}$ and histories consistent with $S$.
$S$ is treated as the problem input (encoded using alphabet $\Sigma$) and reachable histories from $\mathcal{H}_S$ are valid outputs.

Our definition treats the FOSG $G$ and the joint policy $\pi$ as fixed and assumes that the FOSG functions ($\mathcal{T}, \mathcal{A}$, etc.) and the policy can be evaluated in polynomial time with respect to their inputs, and are not encoded as inputs to the problem.
Thus, the input and output sizes of a problem instance are the lengths of the encoded observation sequence and history, respectively.

A relation $R \subset \Sigma^* \times \Sigma^*$ over alphabet $\Sigma$ is \textbf{polynomially balanced} if there exists a polynomial $p$ such that for all $(x,y) \in R, |y| \leq p(|x|)$ (i.e.,  the length of output $y$ is at most polynomial in the input length) (\cite{jerrum1986random}).
$R$ is \textbf{polynomial-time verifiable} if the predicate $(x,y) \in R$ ($xRy$ for short) can be tested in polynomial time. 
Lemma~\ref{lem:balanced} states that this is the case for $\texttt{FILTER}(G, \pi)$ with finite $\mathcal{W}$ and $\mathcal{A}$.
Finite $\mathcal{W}$ and $\mathcal{A}$ imply polynomial balance and polynomial-time verification is done using $|h|$ evaluations of the observation function and policy to check if $S$ is produced and the reach probability is nonzero.
All proofs are in the appendix.

\begin{lemma} \label{lem:balanced}
For any FOSG $G$ with finite $\mathcal{W}, \mathcal{A}$ and arbitrary joint policy $\pi$, $\texttt{FILTER}(G, \pi)$ is polynomially balanced and polynomial-time verifiable.
\end{lemma}

\subsection{Computational Problem Variants}
For a binary relation $R$, there are several naturally associated computational problems. Given a problem instance $x \in \Sigma^*$, some of these are:
\begin{enumerate}
    \item \textbf{Existence}: Is there a $y \in \Sigma^*$ such that $xRy$?
    \item \textbf{Construction}: Return a $y \in \Sigma^*$ such that $xRy$ if one exists.
    \item \textbf{Generation}: Generate a $y \in \Sigma^*$ such that $xRy$ according to some predetermined distribution over the solution set $\{ y\in \Sigma^* : xRy\}$ if one exists.
    \item \textbf{Counting}: Compute $|\{ y \in \Sigma^* : xRy \}|$.
    \item \textbf{Enumeration}: Return all $y \in \Sigma^*$ such that $xRy$.
\end{enumerate}
\nobreak
Of these, generation and enumeration are clearly relevant to history filtering.
Prior work (\cite{schmid2021player,brown2020combining,moravvcik2017deepstack,brown2019superhuman}) has generally relied on enumerative methods, i.e. \ filtering histories by explicitly representing the entire PBS.
Generative methods for history filtering potentially have the advantage of avoiding explicit PBS representation.
In the next section, we analyze the computational complexity of problem variants 1-5.

%% file: sec/filter-complexity.tex
\section{Complexity of Filtering Histories}
Efficient (polynomial-time and space) algorithms for history filtering enable scalable depth-limited search in imperfect information games.
In this section, we provide an FOSG instance where the construction variant of \texttt{FILTER} is intractable and explain when efficient enumeration is possible.

\subsection{FNP-Completeness of Construction}
Consider the following two-player FOSG, based on Functional Boolean Satisfiability (FSAT):
\begin{definition} \label{def:fsat-game}
(\textbf{3-FSAT-GAME}) 
For a given integer $m$, world states are encoded as $m$-variable truth assignments.
Starting at initial state $w^0$, player 1 chooses an action $a^0 = (y_1, y_2, ..., y_m)$ that represents a truth assignment.
This is followed by a transition to some $w = (y_1, y_2, ..., y_m)$ that encodes the same assignment.
$O_{pub}(w^0,a^0,w)$ reveals no public information about the transition to $w$ except that the action and transition occurred.
$w$ has joint action set $\mathcal{A}(w) := \{a\}$ and $\mathcal{T}(w,a) := w$ for all $w \in \mathcal{W}$.
Rewards are arbitrary.
When action $a$ is taken at time $t$, the public observation function $O_{pub}(w,a,w)$ generates, at random, a 3-CNF clause $c_t$ that is satisfied by $w$.
\end{definition}

Player 1 chooses an $m$-variable truth assignment, and then player 2 repeatedly takes arbitrary actions---outputting a 3-CNF clause satisfied by the assignment as a public observation each time.
The truth assignment is unknown to player 2, so solving a 3-CNF Boolean formula is necessary to construct a history consistent with the public observations. 
Additionally, a sequence of observations equivalent to any satisfiable 3-CNF formula over $m$ variables can be generated by playing this game.

\begin{theorem} \label{thm:fnp-complete}
There exists a joint policy $\pi$ for which the construction problem associated with $\texttt{FILTER}(\texttt{3-FSAT-GAME}, \pi)$ is \texttt{FNP}-complete.
\end{theorem}

Theorem~\ref{thm:fnp-complete} implies that unless $\texttt{P} = \texttt{NP}$, computing even a single history corresponding to a given public state in the \texttt{3-FSAT-GAME} is intractable in the worst case.
It follows that the same applies to more complex computations such as generation and enumeration.
However, there are several examples where these computations have been successfully performed in practical examples of games such as poker---we discuss where efficient enumeration is feasible next.

\subsection{Efficient Enumeration in Games with Sparse Public States} \label{sec:efficient-enum}
Prior work has often limited application domains to games where the public state is trivially enumerated and beliefs can be represented explicitly in physical memory; the efficiency of these algorithms depends on a structural property of the game tree that we call \textit{sparsity}.

\begin{definition}
The public tree $\mathcal{S}$ of an FOSG $G$ is \textbf{sparse} if and only if all public states $S = (o^1,o^2,...,o^t)$ in $\mathcal{S}$ satisfy $|\mathcal{H}_S| \leq p(t)$ for some polynomial $p$.
Public trees that do not satisfy this property are \textbf{dense}.
\end{definition}

Public states in games with sparse public trees can be enumerated in polynomial time using a simple breadth-first search that makes at most $|\mathcal{A} \times \mathcal{W}|p(k)$ calls to the observation function at depth $k$ (see proof of Theorem~\ref{thm:sparsity} in the appendix for more details).

\begin{theorem} \label{thm:sparsity}
For any FOSG $G$ with finite $\mathcal{W},\mathcal{A}$, the enumeration problem associated with $\texttt{FILTER}(G, \pi)$ can be solved in polynomial time if and only if $G$'s public tree is sparse.
\end{theorem}

As an example, consider the following variants of two-player poker.
In Texas Hold'em, there is a 52-card deck and two cards are dealt to each player; public states in this game are of constant size.
With $n$ cards in the deck where each player is dealt 2 cards, the number of histories per public state is polynomial in $n$.
Both of these games have sparse public trees.
However, with $n$ cards in the deck and $k$ cards dealt to each player, the number of histories is exponential in $k$, so the public tree is dense.

Sparsity itself does not guarantee that enumeration is feasible on modern hardware.
For instance, physical memory and time constraints still prevent enumeration in large trick-taking card games with constant-sized public states such as Skat and Hearts.
In the next section, we propose an algorithm for history generation in these games and validate it empirically. 

%% file: sec/filter-ttcg.tex
\section{MCMC History Generation in Trick-Taking Card Games}

Trick-taking card games (TTCGs) like Contract Bridge, Skat, and Hearts are played by millions of people worldwide and have historically been of significant interest to the AI community.
However, superhuman-level computer play in these games has yet to be achieved---in part because of their large public belief states.
Here, we devise a Gibbs sampler (\cite{geman1984stochastic}) for history generation in TTCGs that treats histories as states in a Markov chain and uses local computations to generate histories without explicit belief representation.

\subsection{Challenges of Filtering in TTCGs}
Depending on the size of the deck and the number of cards dealt to each player, public states in TTCGs may be large early in the game: up to $n! / (n-k)!$ for decks with $n$ cards and $k$ cards dealt in total.
They shrink as information is revealed because players must be dealt the cards they play, and \textit{void suits} (when a player reveals they cannot have a certain suit through play) imply certain cards cannot be dealt to certain players.
These observations lead to two possible constraints on a card being dealt to a player: either it must have been dealt to a player, or it cannot have been dealt to a player.
Both can be checked efficiently through a single pass over the observation sequence.

\subsection{History Construction in TTCGs}
The solution to history construction for some $S \in \mathcal{S}$ is a history $h \in \mathcal{H}_S$.
This history can serve as an initial state for the Markov process we have designed to solve generation in TTCGs.
As we now describe, construction can be solved in polynomial time using an algorithm for integer maximum flow such as Edmonds-Karp (\cite{edmonds1972theoretical}) along with some simple pre- and post-processing steps.

\begin{figure}[t]
\centering
\includegraphics[width=0.6\linewidth]{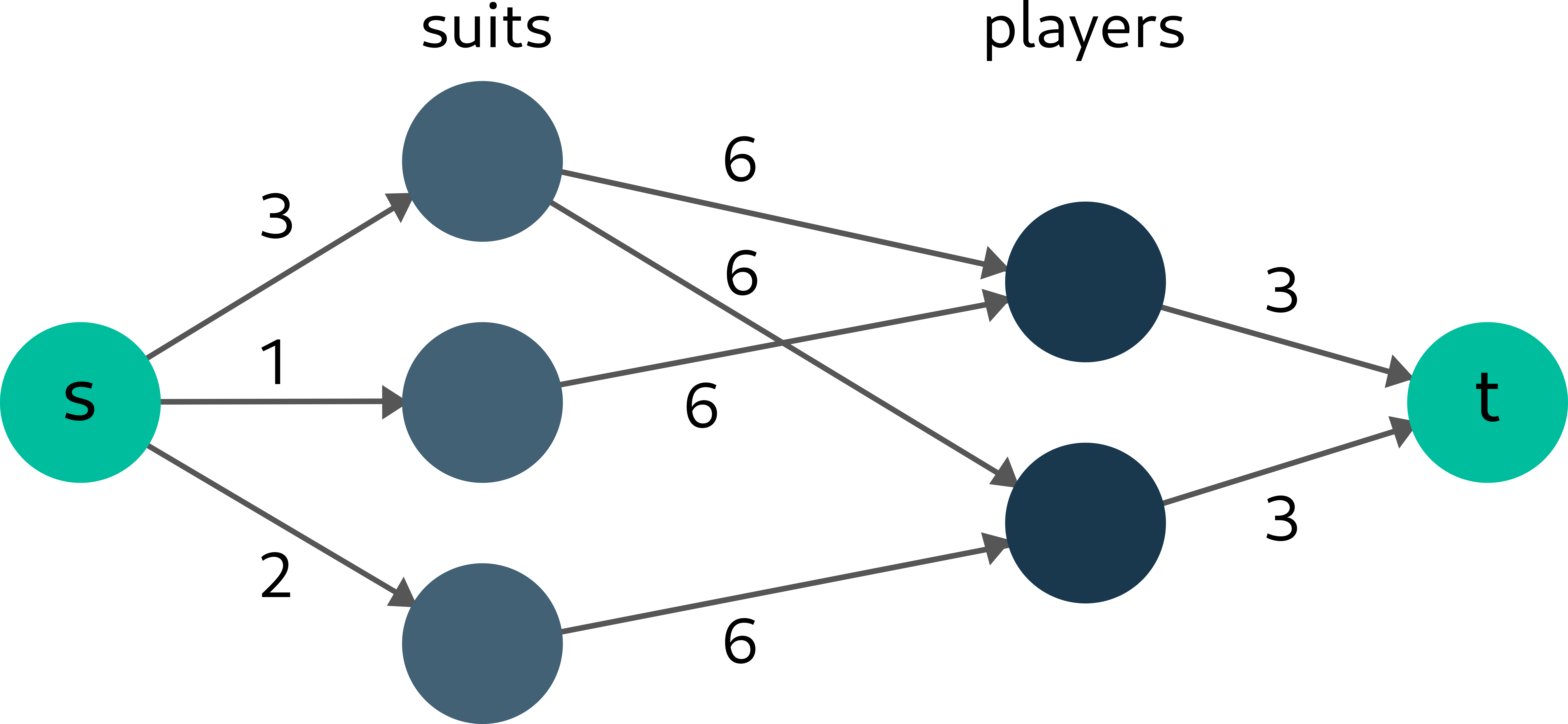}
\caption{A flow network constructed from a history in a TTCG with three suits and two players. The suits have three, one, and two unknown cards remaining, respectively, and each player must be dealt three more cards. The first player is void in the third suit, and the second player is void in the second suit.}
\label{fig:flow-graph}
\end{figure}

Given a history $h$ with the subsequence of private actions that represent the deal, $\sigma \sqsubseteq h$, we create a flow network that captures the constraints of the cards in $\sigma$.
Cards revealed through play must be dealt to the player that played them, so we can ignore them when solving for deals that satisfy the other constraint: where a player cannot have any more cards in a suit.
The source vertex is connected to $k$ \textit{suit} vertices via directed edges with a capacity that corresponds to the number of unknown cards remaining in that suit.
Each suit vertex is connected to a \textit{player} vertex if it is possible for that player to hold cards of that suit in their hand.
The edges connecting the player and suit vertices have capacity equal to the number of total unknown cards remaining in all suits.
Finally, the player vertices are connected to the sink via edges that correspond to the number of cards remaining in that player's hand.
See Figure~\ref{fig:flow-graph} for an example.

\begin{lemma} \label{lem:max-flow}
For TTCG $G$ and joint policy $\pi$ with full support, $\texttt{FILTER}(G, \pi)$ can be solved in polynomial-time using a maximum flow computation.
\end{lemma}

Applying a maximum flow algorithm based on augmenting paths will construct integral flows that represent assignments of the number of unknown cards in each suit dealt to each player---which we call a \textbf{suit length assignment}.
We can select one of possibly many histories that satisfy the suit length assignment (and the cards that have been explicitly revealed through play) and construct a history from $\mathcal{H}_S$.

\subsection{TTCG Gibbs Sampler} \label{sec:gibbs}
The TTCG Gibbs sampler is Markov Chain Monte Carlo method (see \cite{haggstrom2002finite} for an overview) for generative history filtering in TTCGs.
It is based on two concepts: Markov chain states are histories from $\mathcal{H}_S$, and transitions use unnormalized reach probabilities and involve local modifications to the suit length assignment of the current state.

We start by describing the neighbor generation algorithm, \texttt{RingSwap} (Algorithm~\ref{alg:ring-swap}), which operates on suit length assignment matrices.
Constraints on suit length assignments can be encoded using a matrix with row and column sums equal to the number of unknown cards in each suit and player hand.
Void suits are represented as entries fixed to zero.
For example, the following is a suit length assignment matrix that satisfies the max flow network in Figure~\ref{fig:flow-graph}:
\nobreak
\begin{equation*}
\begin{bmatrix}
2 & 1 & 0\\
1 & 0 & 2
\end{bmatrix}
\end{equation*}
The rows sum to 3 because each player has 3 unknown cards, and the columns sum to the number of unknown cards in the corresponding suit.

\begin{algorithm}[t]
\SetAlgoLined
\DontPrintSemicolon
\SetKwInOut{Input}{input}\SetKwInOut{Output}{output}
\Input{$S$ --- public state, $\sigma$ --- deal consistent with $S$}
\Output{$\Omega_\sigma$ --- set of neighbors of $\sigma$}
\textbf{let} $A_{n \times m}$ be the suit assignment matrix for $\sigma$ \;
$\Omega_\sigma \leftarrow \{\}$ \;
\For{Row $i$ in rows($A$)} {
    \For{Columns $j, k, j \not = k$ in cols($A$)}{
        \textbf{if} void($S, i, j$) or void($S, i, k$) or $a_{i,k} = 0$, \textbf{continue} \;
        $C \leftarrow A$ \;
        $c_{i,j} \leftarrow c_{i,j} + 1; c_{i,k} \leftarrow c_{i,k} - 1$ \;
        $\Omega_\sigma \leftarrow \Omega_\sigma \bigcup$ \texttt{BFS}$(C, n)$ to find all ways to make $C$ a valid suit length assignment
    }
}
\Return $\Omega_\sigma$
\caption{RingSwap\label{alg:ring-swap}}
\end{algorithm}

With suit assignment matrix $A$ corresponding to $\sigma$, \texttt{RingSwap} repeats the following for all players $i$.
For every pair of non-void suits $j$ and $k$, perform a \textit{swap} by adding a card to $A_{i,j}$ and removing one from $A_{i,k}$.
The column sums of the matrix are now incorrect ($j$ has too many cards and $k$ has one too few), and must be corrected via a sequence of swaps in other rows.
All sequences of swaps of length $< n$ that lead to valid suit length assignments are then computed via BFS and a valid assignment is selected proportionally to the number of histories it corresponds to.

We can now describe the \textit{TTCG Gibbs sampler}.
For public state $S$ at time $t$, given $X_t = h$, deal $\sigma \sqsubseteq h$, and a joint policy $\pi$ with full support at all infostates, consider the following Markov chain:

At time $t+1$:
\begin{enumerate}
\item Compute $\Omega_\sigma$, the set of all neighbors of $\sigma$ using the procedure \texttt{RingSwap}$(S, \sigma)$
\item Sample $\sigma^\prime$ uniformly from $\Omega_\sigma$
\item Compute $\Omega_{\sigma^\prime}$ and $h^\prime$ such that $\sigma^\prime \sqsubseteq h^\prime$ by replacing $\sigma$ with $\sigma^\prime$ in $h$ to form $h^\prime$
\item Let $z = \min \{  1, \frac{\bar{P^\pi}(h^\prime) |\Omega_\sigma|}{\bar{P^\pi}(h) |\Omega_{\sigma^\prime}|} \}$
\item With probability $z$, $X_{t+1} = h^\prime$, otherwise $X_{t+1} = h$
\end{enumerate}

State transitions are done according to the Metropolis-Hastings algorithm (\cite{metropolis1953equation,hastings1970monte})--- with unnormalized reach probabilities $\bar{P}^\pi$ as $\mu^*$ and uniform selection over the neighboring states.
All computations are local to the current history at time $t$, and take at most polynomial time in the history length.
The following theoretical details state that the chain is suitable for unbiased history generation.

\begin{theorem} \label{thm:ttc-irreducible}
The TTCG Gibbs sampler is aperiodic and irreducible.
\end{theorem}

Theorem~\ref{thm:ttc-irreducible} implies that the TTCG Gibbs Sampler converges to some stationary distribution;
the following theorem ensures that its stationary distribution is the desired $P^\pi$.

\begin{theorem} \label{thm:ttcg-reversibility}
The stationary distribution of the TTCG Gibbs sampler with input $\pi$ is $P^\pi$.
\end{theorem}

Given an initial history obtained by solving the construction problem for the game and a policy with full support at all infostates, the TTCG Gibbs sampler correctly generates histories from $P^\pi(\cdot|S)$ in the limit.
The next section validates the efficiency and approximation quality of the TTCG Gibbs sampler empirically in the domain of Oh Hell (\cite{parlett2008penguin}); we leave the theoretical analysis of its mixing time to future work.

%% file: sec/experiments.tex
\begin{figure*}[t]
     \centering
     \begin{subfigure}{0.32\linewidth}
         \centering
         \includegraphics[width=\linewidth]{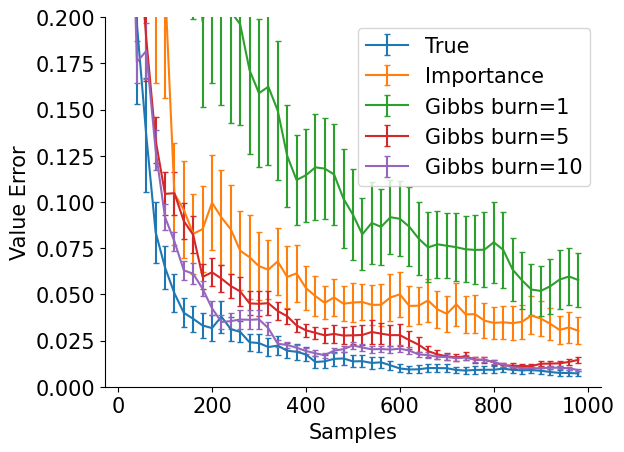}
         \caption{192 Histories}
         \label{fig:value-small}
     \end{subfigure}
     \hfill
     \begin{subfigure}{0.32\linewidth}
         \centering
         \includegraphics[width=\linewidth]{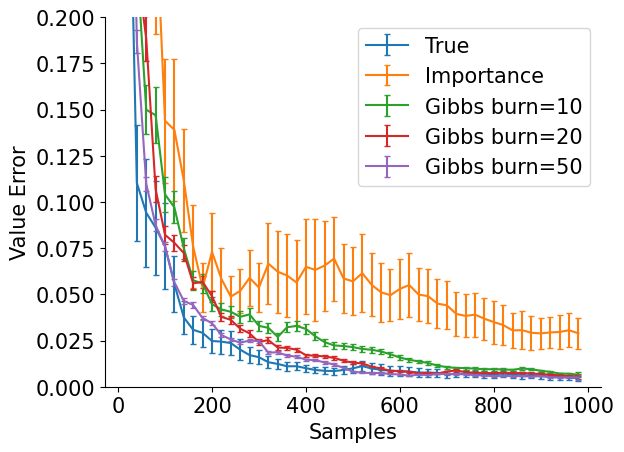}
         \caption{12,960 Histories}
         \label{fig:value-med}
     \end{subfigure}
     \hfill
     \begin{subfigure}{0.32\linewidth}
         \centering
         \includegraphics[width=\linewidth]{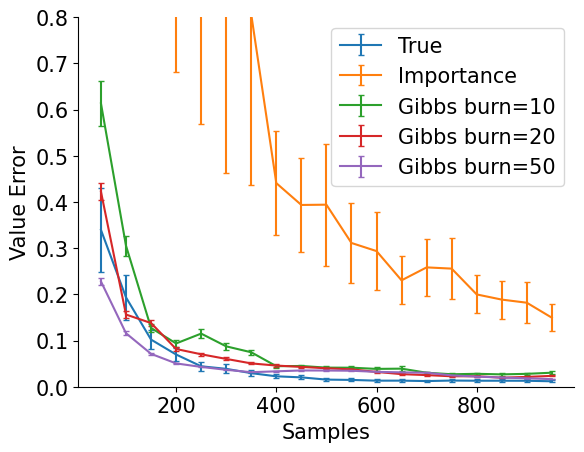}
         \caption{544,320 Histories}
         \label{fig:value-large}
     \end{subfigure}
    \caption{Value estimation error of TTCG Gibbs Sampler with specified burn-in and baselines on PBS of various sizes. Error bars show one standard error of the mean over 100 runs. See the appendix for full game parameters.}
    \label{fig:value-curve}
\end{figure*}

\begin{figure*}
     \centering
     \begin{subfigure}{0.32\linewidth}
         \centering
         \includegraphics[width=\linewidth]{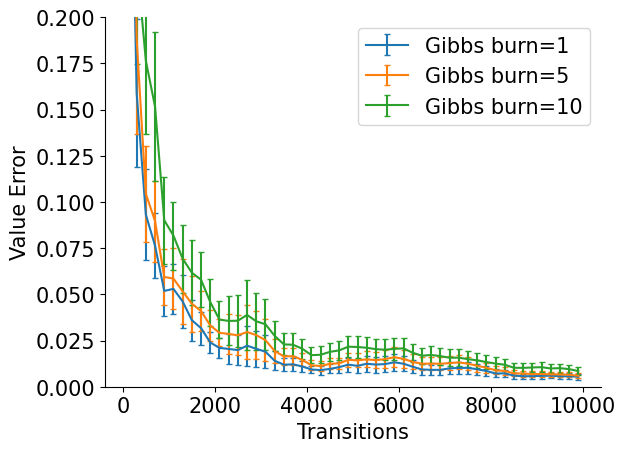}
         \caption{192 Histories}
         \label{fig:mixing-small}
     \end{subfigure}
     \hfill
     \begin{subfigure}{0.32\linewidth}
         \centering
         \includegraphics[width=\linewidth]{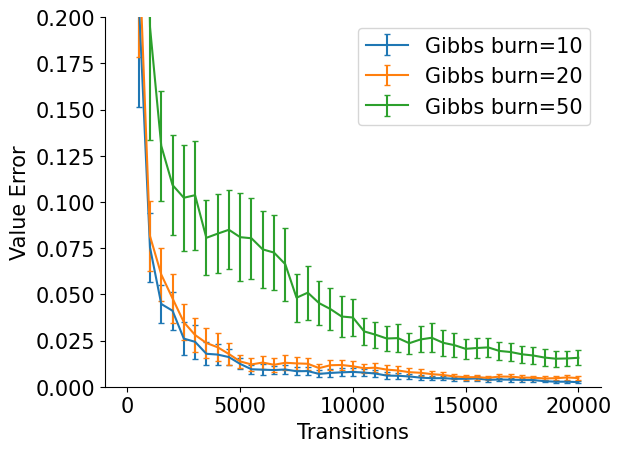}
         \caption{12,960 Histories}
         \label{fig:mixing-med}
     \end{subfigure}
     \hfill
     \begin{subfigure}{0.32\linewidth}
         \centering
         \includegraphics[width=\linewidth]{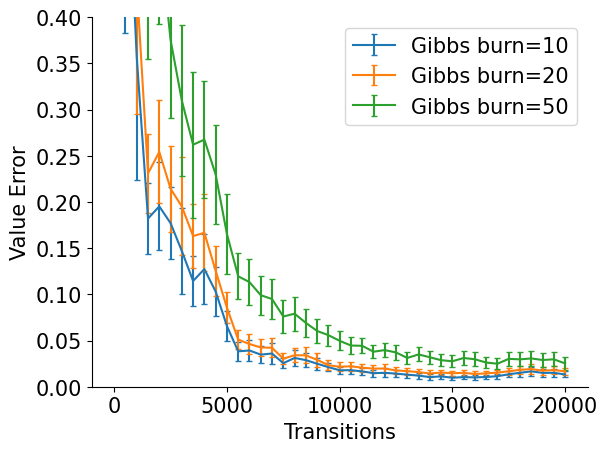}
         \caption{544,320 Histories}
         \label{fig:mixing-large}
     \end{subfigure}
    \caption{The effect of the number of samples burned on the value estimation error of the TTCG Gibbs Sampler. Burning fewer samples can result in a lower value error after an equal number of transitions. Error bars show one standard error of the mean over 100 runs.}
    \label{fig:mixing-time}
    \vspace{-5mm}
\end{figure*}

\subsection{Experiments}
The TTCG Gibbs sampler we just described runs in polynomial-time with respect to history length and removes dependencies on explicit belief representation.
Thus, the algorithm's scalability depends on the mixing time of the underlying Markov chain.
The following empirical results suggest that the chains mix rapidly in this domain.

We evaluate the TTCG Gibbs sampler using a value estimation task in the trick-taking card game Oh Hell.
Oh Hell is an $N$-player game with $n$ cards in the deck, where each player is dealt between $1$ and $\lfloor (n-1)/N \rfloor$ cards.
We control the size of randomly generated $\mathcal{H}_S$---starting small and then scaling to $\mathcal{H}_S$ several orders of magnitude larger---by varying the number of cards in the deck and the number of cards dealt to players.
The task is to estimate the expected value of $S$ under $\pi$, $V^\pi(S) = \sum_{h \in S}P^\pi(h)V^\pi(h)$,
where the value of a history $h$, $V^\pi(h) = \sum_{z \in \mathcal{Z}: h \sqsubseteq z} P^\pi(h,z) u(z)$ is the expected utility of the terminal histories in $\mathcal{Z}$ reachable from $h$.
The initial state is chosen uniformly from $\mathcal{H}_S$, and ranges are constructed using policies learned via independent Q-learning (\cite{claus1998dynamics})---hyperparameters and other details can be found in the supplementary material.
Strong performance in this task does not necessarily imply improved search performance; instead, it demonstrates the sampler's ability to generate histories from the correct public belief state.

Figure~\ref{fig:value-curve} shows the value error curves of the TTCG Gibbs sampler with a specified burn-in compared to two Monte Carlo baselines.
\texttt{True} samples from the exact joint range, and \texttt{Importance} performs (biased) importance sampling with a uniform proposal and unnormalized reach probabilities.
The latter estimates the PBS value by correcting the weight of samples drawn uniformly from $\mathcal{H}_S$.
In practice---especially at larger scales than these experiments---importance sampling may not be feasible without a generation algorithm that approximates a known proposal distribution in polynomial time.
We see that the TTCG Gibbs sampler outperforms importance sampling and closely approximates sampling from the joint range in all cases while using a burn-in orders of magnitude smaller than the size of the public belief states.

Figure~\ref{fig:mixing-time} shows the effect of burning samples on value estimation error.
On one hand, our results show that burning fewer samples can result in a better value estimate with fewer state transitions.
This is unsurprising because larger sample sizes generally lead to better estimates. 
However, evaluating a sample could be several orders of magnitude more costly than performing a state transition in the Markov chain in practice.
Comparing Figures~\ref{fig:mixing-time}~and~\ref{fig:value-curve} shows that, if the estimation task is constrained by time or computational resources, fewer high-quality samples may produce a better estimate.
The appropriate burn-in depends both on the task and the available resources.

Our experiments show that the TTCG Gibbs Sampler leads to increased scalability over the enumerative approach in this value estimation task.
With a small burn-in of 20 state transitions, a good approximation of the PBS value is achieved after only 400 samples and therefore only 8,000 total state transitions---compared to the enumerative approach which must construct and calculate the reach probability for 544,320 histories.
State transitions are computed locally to the current state (history) and do not require knowledge of the rest of the PBS, so memory requirements are dramatically reduced.
These effects increase along with public state size; in Oh Hell with a 52-card deck, public states can have over $10^{61}$ histories.

%% file: sec/concl.tex
\section{Discussion and Conclusions}
In this paper, we analyzed the computational complexity of history filtering for subgame decomposition and depth-limited search in imperfect information games.
Although even the simplest form of the computation is likely intractable in general, we have shown that depth-limited search remains a viable option in certain classes of imperfect information games.
Efficient enumeration is achievable in games with polynomially-sized public states; many application domains from prior work seem to have this property. 
However, generative methods for history filtering may not require explicit belief representation, and are therefore more scalable than enumeration.
To this end, we have introduced a novel generation algorithm for trick-taking card games for asymptotically correct and efficient history filtering that fits seamlessly with Monte Carlo-style search algorithms.

Our TTCG Gibbs sampler needs a method that constructs a valid history from the public state but is otherwise flexible.
It does not depend on knowing the size of the public state and does not require normalized reach probabilities to
produce samples with the correct probability.
Unlike the enumerative approach which front-loads all of its computation, our algorithm applies to the setting where the game-playing algorithm must return an action within a predefined time budget.
It is also easy to parallelize; multiple samples can be generated simultaneously by starting the process multiple times from the same initial state and running each for the desired burn-in time.
We demonstrate its effectiveness empirically---though future work should prove rapid mixing analytically.

\section*{Acknowledgements}
This work was funded by the Canada CIFAR AI Chairs Program.
We acknowledge the support of the Natural Sciences and Engineering Research Council of Canada (NSERC).

%% file: sec/extra.tex
\section*{Alternative Definitions of Filtering}
We analyze history filtering in terms of the public state and joint range for simplicity, but some algorithms compute value functions using player ranges and beliefs instead.
Definitions~\ref{def:filter-belief}~and~\ref{def:filter-range} alternatively describe filtering histories from a player's infostate according to their beliefs and filtering a player's infostates from the public state according to their range, respectively.

\begin{definition} \textbf{(FILTER-BELIEF)} \label{def:filter-belief}
For any FOSG $G$ with finite $\mathcal{W}, \mathcal{A}$ and joint policy $\pi$, let $\texttt{FILTER-BELIEF}(G, \pi) := \{(S_i, h) \in \Sigma^* \times \Sigma^* : s_{i}(h) = S_i, P^\pi(h|S_i) > 0\}$.
\end{definition}

\begin{definition} \textbf{(FILTER-RANGE)} \label{def:filter-range}
For any FOSG $G$ with finite $\mathcal{W}, \mathcal{A}$ and joint policy $\pi$, let $\texttt{FILTER-RANGE}(G, \pi) := \{(S, S_i) \in \Sigma^* \times \Sigma^* : S_i \in \mathcal{S}_i(S), P^\pi(S_i|S) > 0\}$.
\end{definition}

The key difference in Definition~\ref{def:filter-belief} is that the input is a player $i$'s infostate $S_i$ instead of a public state $S$.
Definition~\ref{def:filter-range} consists of public state $S$ as input and $i$'s infostates as output.
Both require that the output history is reachable (with non-zero probability) according to the joint policy $\pi$.

\begin{lemma} \label{lem:belief-reduction}
\texttt{FILTER-BELIEF} and \texttt{FILTER} are mutually polynomial-time reducible.
\begin{proof}
Given an instance $S$ of $\texttt{FILTER-BELIEF}(G, \pi)$ with FOSG $G$ and player $i$ with infostate tree $\mathcal{S}_i$ in $G$, there exists an FOSG $G^\prime$ with public tree $\mathcal{S}^\prime = \mathcal{S}_i$.
Given $G$, we need to construct $G^\prime$ in polynomial-time with respect to the encoding length for the input instance $S$, using only the functions that define $G$.
To do so, we include evaluations of $i$'s private observation function $s_i$ in the public observation function $s^\prime_{pub}$ of $G^\prime$.
In other words, we set $s^\prime_{pub} = (s_i, s_{pub})$ in $G^\prime$, which increases the cost of evaluating $s^\prime_{pub}(h)$ by a factor of 2.
Since each call to the observation function is polynomial-time with respect to the encoding of $w,a, w^\prime$, our transformation is also polynomial-time.

A basic fact from this construction is $h \in S_i \iff h \in S^\prime$.
Moreover, there exists a joint policy $\pi^\prime$ such that $P^{\pi^\prime}(h) = P^\pi(h)$ for each such $h$.
Joint policy $\pi^\prime$ is constructed by mapping $\pi^\prime_p((s^\prime_p(g), s^\prime_{pub}(g)),a) = \pi_p((s_p(g), s_{pub}(g)), a)$ for all $g \cdot a \sqsubseteq h$  for each player $p \in \mathcal{N}$.
Each player $p \neq i \in G^\prime$ ignores $i$'s private information, reconstructing an information set on which $\pi_p$ is valid and contributes the desired reach probability on $h$.
Evaluating $\pi^\prime$ therefore requires just a single call to $\pi$, so the transformation from $\pi$ to $\pi^\prime$ is clearly polynomial-time with respect to the encoding length of $S$. 

Since $S = S^\prime$, we have $P^\pi(h|S_i) = P^{\pi^\prime}(h|S^\prime)$.
It follows that $(S_i,h) \in \texttt{FILTER-BELIEF}(G, \pi) \iff (S^\prime, h) \in \texttt{FILTER}(G^\prime, \pi^\prime)$, where $G^\prime$ and $\pi^\prime$ can be constructed in polynomial time given $G$ and $\pi$.

For the other direction, given an instance of $\texttt{FILTER}(G, \pi)$ with FOSG $G$, construct $G^\prime$ by adding a player $i$ (who takes no actions) and setting $s_i^\prime = s_{pub}$, $s^\prime_{pub} = \emptyset$, and $s^\prime_{j} = (s_{pub}, s_j)$ for every $j \in -i$, and similarly constructing a new joint policy $\pi^\prime$ that copies $\pi$ at every infostate for players in $-i$.
It follows that $(S,h) \in \texttt{FILTER}(G, \pi) \iff (S_i^\prime, h) \in \texttt{FILTER-BELIEF}(G^\prime, \pi^\prime)$.
\end{proof}
\end{lemma}

\begin{lemma} \label{lem:range-reduction}
\texttt{FILTER} $\leq_p$ \texttt{FILTER-RANGE}.
\begin{proof}
Given an instance $S$ of $\texttt{FILTER}(G, \pi)$, we construct game $G^\prime$ by adding a player $i$ that has perfect information but takes no actions.
This transformation is computed in polynomial-time with respect to the encoding length of $S$ by creating an observation function that returns the current history to $i$.
Thus, for any $S \in \mathcal{S}$ in the original game $G$, there exists an $S_i^\prime$ in $i$'s infostate tree which corresponds to exactly one $h \in S$.
Since $i$ takes no actions, the policy is unchanged.
This means we have $P^\pi(h|S) = P^\pi(S_i^\prime|S)$ by the definition of infostate reach probability.
Therefore, $(S, S_i^\prime) \in \texttt{FILTER-RANGE}(G^\prime, \pi) \implies (S,h) \in \texttt{FILTER}(G, \pi)$.
\end{proof}
\end{lemma}

Lemmas~\ref{lem:belief-reduction}~and~\ref{lem:range-reduction} show that \texttt{FILTER} is polynomial-time reducible to these alternative definitions, and unsurprisingly, that filtering a history from a player's beliefs is equivalent to filtering a history from the joint range.

\section*{Experiment Parameters}

Table~\ref{tab:exp-params} shows the game parameters used to generate Figures~\ref{fig:value-curve}~and~\ref{fig:value-curve-random}.
Experiments are repeated 100 times and runtimes range from seconds to 12 hours on a single core of an AMD Ryzen 9 \copyright \ CPU.

\begin{table}[h]
\small
\centering
\caption{Value estimation and mixing time experiment parameters.}
\label{tab:exp-params}
\begin{tabular}{r c c c c c}\toprule
Size & Players & Suits & Ranks & Tricks & Tricks Played \\
\midrule
192 & 3 & 2 & 4 & 2 & 1 \\
12,960 & 3 & 3 & 4 & 3 & 2 \\
544,320 & 3 & 3 & 4 & 3 & 1 \\
\bottomrule
\\
\end{tabular}
\end{table}

\section*{Additional Experiments}
\begin{figure*}[t]
     \centering
     \begin{subfigure}{0.32\linewidth}
         \centering
         \includegraphics[width=\linewidth]{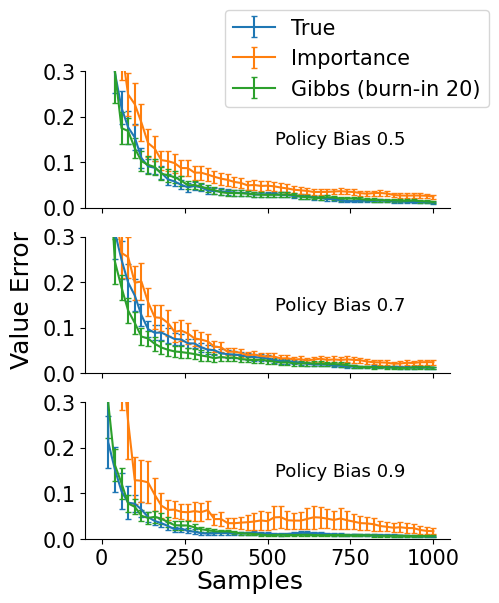}
         \caption{192 Histories}
         \label{fig:value-small-random}
     \end{subfigure}
     \hfill
     \begin{subfigure}{0.32\linewidth}
         \centering
         \includegraphics[width=\linewidth]{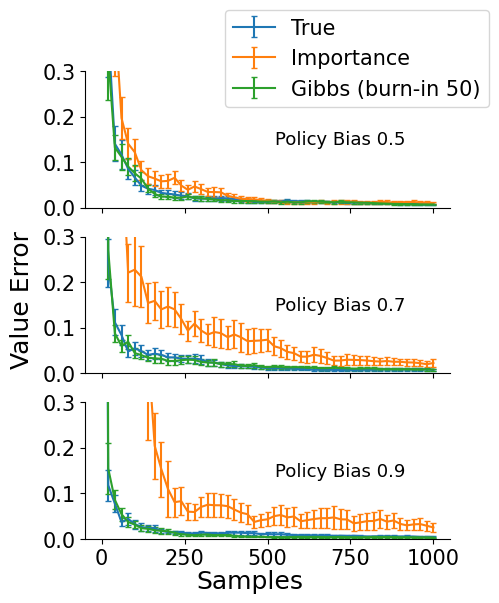}
         \caption{12,960 Histories}
         \label{fig:value-med-random}
     \end{subfigure}
     \hfill
     \begin{subfigure}{0.32\linewidth}
         \centering
         \includegraphics[width=\linewidth]{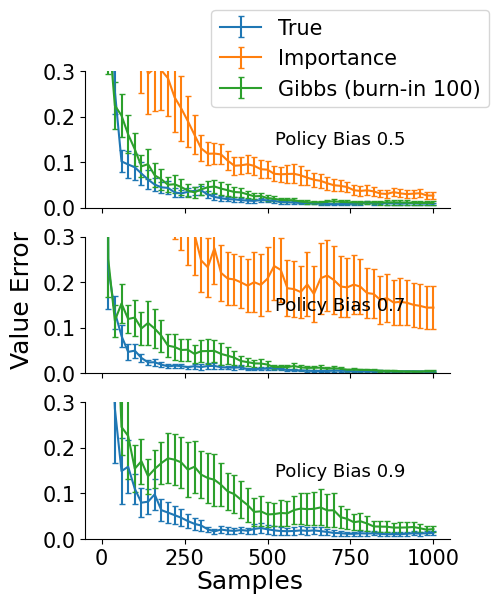}
         \caption{544,320 Histories}
         \label{fig:value-large-random}
     \end{subfigure}
    \caption{Value estimation error of TTCG Gibbs Sampler with specified burn-in and baselines on PBS of various sizes. Error bars show one standard error of the mean over 100 runs. Joint policies are generated randomly according to \textit{policy bias}.}
    \label{fig:value-curve-random}
\end{figure*}

In addition to the value estimation experiments using policies learned via reinforcement learning in the main paper, we conducted similar experiments using random policies to increase empirical coverage of the policy space.
At any information state, the policies place \textit{policy bias} probability on a randomly selected action and distribute the remaining probability mass uniformly across the other actions.
Figure~\ref{fig:value-curve-random} shows that these experiments yielded similar results to the RL policies.

Table~\ref{tab:exp-stats} shows the entropy and variance of the tested public states, organized by the policy bias used to generate them.
Policy bias clearly has an effect on both the reach probabilities of the histories in the PBS, as well as on the variance of the PBS value.
The medium size exhibits proportionally higher entropy and lower variance because an extra trick has been played---meaning play is closer to the end of the game.

\begin{table}[h]
\small
\centering
\caption{Mean variance and entropy of generated public belief states of different sizes.}
\label{tab:exp-stats}
\begin{tabular}{lcccccl}\toprule
& \multicolumn{3}{c}{Entropy} & \multicolumn{3}{c}{Variance}
\\\cmidrule(lr){2-4}\cmidrule(lr){5-7}
 Bias      & 0.5  & 0.7  & 0.9  & 0.5  & 0.7 & 0.9\\
 \midrule
196  & $6.84\pm0.05$ & $6.13\pm0.07$ & $4.89\pm0.07$ & $12.98\pm1.12$ & $10.11\pm0.95$ & $5.36\pm0.93$ \\
12,960 & $12.04\pm0.06$ & $10.99\pm0.08$ & $9.30\pm0.08$ & $8.77\pm1.02$ & $6.01\pm0.86$ & $2.38\pm0.58$ \\
544,320 & $16.21\pm0.25$ & $14.06\pm0.45$ & $12.36\pm0.17$ & $10.55\pm1.17$ & $11.44\pm2.40$ & $8.77\pm2.01$ \\
\bottomrule
\\
\end{tabular}
\end{table}

%% file: sec/proofs.tex
\section*{Proofs of Theorems}

\textbf{Lemma~\ref{lem:balanced}.}
For any FOSG $G$ with finite $\mathcal{W}, \mathcal{A}$ and arbitrary joint policy $\pi$, $\texttt{FILTER}(G, \pi)$ is polynomially balanced and polynomial-time verifiable.

\begin{proof}
We need to show that there exists a polynomial $p$ such that for any $(S,h) \in \texttt{FILTER}(G, \pi)$,  $|h| \leq p(|S|)$ and that the predicate $(S,h) \in \texttt{FILTER}(G, \pi)$ can be verified in polynomial time.
$\mathcal{W}$ being finite implies there exists an encoding for all $w \in \mathcal{W}$ and a constant $c_\mathcal{W}$, where $|w| \leq c_\mathcal{W}$.
Likewise, $\mathcal{A}$ being finite implies the existence of an encoding for all $a \in \mathcal{A}$ and a constant $c_\mathcal{A}$ such that $|a| \leq c_\mathcal{A}$.
$h = (w^0,a^0,w^1,a^1,...,w^t)$, so $|h| \leq t(c_\mathcal{W} + c_\mathcal{A}) + c_\mathcal{W}  \leq ct$ for all $t > 0$ and some constant $c > 0$.
For any encoding of input observations $S := (o^1,o^2,...,o^t)$ with  $|o^i| \geq 1$, $|S| \geq t$, so $|h| \leq c|S|$.
Lastly, predicate $(S,h) \in \texttt{FILTER}(G, \pi)$ is easily verified in polynomial time in the encoding length of $S$ and $h$ by checking that $O(w^k,a^k,w^{k+1}) = o^{k+1}$ for $0 \leq k \leq t-1$ and $P^\pi(h) > 0$.
\end{proof}

\textbf{Theorem~\ref{thm:fnp-complete}.}
There exists a joint policy $\pi$ for which the construction problem associated with $\texttt{FILTER}(\texttt{3-FSAT-GAME}, \pi)$ is \texttt{FNP}-complete.

\begin{proof}
A construction problem is in the class \texttt{FNP} if its associated relation is both polynomially-balanced and polynomial-time verifiable (\cite{bellare1994complexity}).
A construction problem is \texttt{FNP}-complete if and only if it belongs to \texttt{FNP} and all other problems in \texttt{FNP} are polynomial-time reducible to it.

First, Lemma~\ref{lem:balanced} implies that the problem is in \texttt{FNP}.
Next, we need to show that \texttt{3-FSAT-GAME}, an \texttt{FNP}-complete problem, can be reduced to $\texttt{FILTER}(\texttt{3-FSAT-GAME}, \pi)$.

Given TM $M$, which computes construction for $\texttt{FILTER}(\texttt{3-FSAT-GAME}, \pi)$ and any \texttt{3-FSAT} instance $\phi = (c_1, ..., c_k)$ containing variables $y_1,...,y_m$ in $\texttt{DTIME}(p(|x|))$ for some time-constructible polynomial $p$, 
map $\phi$ to the equivalent public state $x = \phi$ in \textit{3-FSAT-GAME} and run $h = M^\prime(x)$, where $M^\prime$ simulates $M$ with a time bound of $p(|x|)$ in $DTIME(p(|x|)^2)$.

Let $\pi_1(w^0)$ be the uniform policy, which assigns equal action probability to each of the $2^m$ actions available at $w^0$
Since action $a$ is the singular joint action at any $w \neq w^0$, $\pi_i(s, a) = 1$ for any $s \in S_i(s_{pub})$ for all players $i$---implying that any $h$ output by $M^\prime$ satisfies $P^\pi(h) > 0$. 
If $M^\prime$ returns $h = (w^0,a^0,w,a^1,w,a^2,...,w)$, then $w$ must be a satisfying assignment to $\phi$.
If $M^\prime$ returns \texttt{NO}, then there is no $h, s_{pub}(h) = S$ and $\phi$ is unsatisfiable.
\end{proof}

\textbf{Theorem~\ref{thm:sparsity}.}
For any FOSG $G$ with finite $\mathcal{W},\mathcal{A}$, the enumeration problem associated with $\texttt{FILTER}(G, \pi)$ can be solved in polynomial time if and only if $G$'s public tree is sparse.
\begin{proof}
Suppose $G$'s public tree is sparse. 
To enumerate $\mathcal{H}_S$, consider a basic breadth-first search which, at time $k$, only searches actions that satisfy $O_{pub}(w^k,a^k,w^{k+1}) = O^{k+1}$, where $a^k \in \mathcal{A}(w^k)$ and $w^{k+1}$ is in the support of $\mathcal{T}(w^k,a^k)$.
Let the set $\mathcal{H}^k_S = \{h^\prime \in \mathcal{H} : h^\prime \sqsubseteq h, h \in \mathcal{H}_S, |h^\prime| = k\}$ contain the length $k$ prefix histories of all histories in $\mathcal{H}_S$.
All histories in $\mathcal{H}^k_S$ must produce the same public observation sequence and therefore must correspond to a unique public state.
Since $G$'s public tree is sparse, this implies there can be at most $p(k)$ histories at depth $k$ of the search.
This means we must evaluate $O_{pub}(\cdot)$ at most $|\mathcal{A} \times \mathcal{W}|p(k)$ times at depth $k$ and therefore $O(|h|p(|h|))$ times overall. 

Conversely, suppose there exists a polynomial-time TM $M$ for solving the enumeration problem associated with $\texttt{FILTER}(G, \pi)$.
Assume there exists a dense public state $S$ in $G$.
Then for input instance $S$, $M(S)$ computes $\mathcal{H}_S$ in $p(t)$-time for some polynomial $p$.
This is impossible since, for any polynomial $p^\prime$, $|\mathcal{H}_S| > p^\prime(t)$ by assumption.
\end{proof}

\textbf{Lemma~\ref{lem:max-flow}}
For TTCG $G$ and joint policy $\pi$ with full support, $\texttt{FILTER}(G, \pi)$ can be solved in polynomial-time using a maximum flow computation.

\begin{proof}
Given public state instance $S$, let $N = (V,E)$ be the flow network constructed as follows.
The network contains a source connected via edges to a set of \textit{suit} vertices (one vertex for every suit, each with an edge from the source).
Each suit vertex is connected to any number of vertices from the set of \textit{player} vertices (one for each player).
A suit vertex is connected via a directed edge to a player vertex if and only if the player can hold that suit in their hand.
Finally, each suit vertex is connected to the sink via a directed edge.
For the edges connected to source $s$, capacity $c(s,u)$ is the number of cards remaining in suit $u$.
Likewise, for edges connected to the sink $t$, $c(v,t)$ is the number of cards that must be assigned to player $v$.
Edges connecting suits to players have infinite capacity.

Let $\mathcal{F}$ be the set of flows over $N$, and $\phi : \mathcal{H} \rightarrow \mathcal{F}$ be the following transformation from history to flow: for edges from source to suit and player to sink, the flow is equal to the capacity, for edges from suit to player, the flow is equal to the number of cards from that suit dealt to that player in $h$.

Suppose there exists some $h \in \mathcal{H}_S$, then by construction $|\phi(h)| = \sum_u c(u,t) = \sum_u c(s,u)$ and is an integer because the number of cards dealt to each player is known. 
To show that $|\phi(h)|$ is a maximum flow, assume for contradiction, that there exists some $f \in \mathcal{F}$ such that $|f| > |\phi(h)|$.
This implies that $|f| > \sum_u c(s,u)$ which is impossible, so $|\phi(h)| \geq |f|$ for all $f \in \mathcal{F}$.

Now suppose we have some maximum flow $f^* \in \mathcal{F}$; we can construct $h \in \mathcal{H}_S$ in the following way. 
If $|f*| < \sum_u c(s,u)$, then at least one player or suit cannot have the appropriate number of cards allocated to it, so we return that $\mathcal{H}_S$ is empty.
Otherwise, for each suit $u$ and player $v$, we assign $f^*(u,v)$ arbitrary cards (which have not been revealed by play) from suit $u$ to player $v$ in the deal.
\end{proof}

\textbf{Theorem~\ref{thm:ttc-irreducible}.}
The TTCG Gibbs sampler is aperiodic and irreducible.

\begin{proof}
Self-transitions in the chain imply aperiodicity.
We prove irreducibility by showing that there exists a path between any two consistent suit length assignments that can be generated via multiple iterations of our neighbor generation algorithm.
This proves the chain's irreducibility because the neighbor set is the union of all history subsets that correspond to neighboring suit length assignments.

Given two suit length assignment matrices $A$ and $B$ (see description in Section~\ref{sec:gibbs}, we prove that $B$ is reachable from $A$ by showing that the $L_1$-norm
\nobreak
\begin{equation*}
    ||B-A||_1 = \sum_{j=1}^n\sum_{i=1}^m|b_{i,j} - a_{i,j}| 
\end{equation*}
decreases to zero for a sequence of iterations of our neighbor generation algorithm, and that each iteration makes at most $\max\{n,m\}$ \textit{swaps} (where a player gains a card of one suit and loses a card of another) for an $n \times m$ matrix.

First, we show that applying a sequence of \texttt{RingSwap} calls to $A$ decreases $||B-A||_1$ to zero.
Let $C = B-A$, and denote a swap that adds one to $c_{i,j}$ and subtracts one from $c_{i,k}$ as $\delta = (i;j,k)$.
Since all columns and rows of $C$ sum to zero, if $||C||_1 > 0$, there must be some element $c_{i,j} < 0$ and another $c_{i,k} > 0$.
Since swap $\delta_0 = (i;j,k)$ decreases $||C||_1$ by 2, and such a swap is guaranteed to exist whenever $||C||_1 > 0$, repeating this process decreases $||C||_1$ to zero.

After applying $\delta_0$, $\sum_b C_{a,b} = 0$ for all rows $a$, but columns $\sum_aC_{a,k} = -1$ and $\sum_aC_{a,j} = 1$, so the result does not correspond to a valid suit assignment.
To generate a neighbor, the algorithm performs a sequence of swaps until the column sums are corrected.
Since, prior to $\delta_0$, $c_{i,k} > 0$, there exists $c_{l,k} < 0$; likewise there must be some $c_{l,z} > 0$.
Perform swap $(l;k,z)$, now $\sum_aC_{a,k} = 0$.
If $z=j$, then  $\sum_aC_{a,j} = 0$ and we are done.
Otherwise, $\sum_aC_{a,z} = 1$, and the above process can be repeated using column $z$ instead of $l$.

Now, we show that given any starting swap, we can reach a valid suit length assignment using at most $n$ swaps.
Assume $n \geq m$, otherwise transpose $A$ and $B$.
As we have shown above, given an arbitrary starting swap $(i;a,b)$, all rows and columns will sum to zero once a swap of the form $(j;b,c)$ adds a unit back to column $b$.
Consider the sequence of swaps made until the $z=j$ stopping condition is reached and suppose that a row has to be repeated in this sequence (i.e. there is a subsequence $(k;d,e)$,..., $(k;f,g)$).
This means the remainder of the sequence begins by with a swap from column $g$ and ends with a swap into column $b$, and leads to the condition that all rows and columns sum to zero.
But before the first swap in the subsequence, $(k;d,e)$, only columns $d$ and $b$ have non-zero sum.
This implies we can remove this subsequence, and replace it with $(k;d,g)$ and proceed until $b$ is reached.
So, each row must only be visited once in the sequence of swaps, implying that there exists a sequence with length at most $n$ which turns $A$ to $A^ \prime$ where all rows and columns of $A^\prime - B$ sum to zero and $||A^\prime - B||_1 < ||A - B||$.
This implies a path between any two suit length assignments can be generated using multiple ring swaps of length $<=n$.
\end{proof}

\textbf{Theorem~\ref{thm:ttcg-reversibility}.}
The stationary distribution of the TTCG Gibbs sampler with input $\pi$ is $P^\pi$.

\begin{proof}
A sufficient condition for some $\mu$ to be the stationary distribution of a Markov chain is that the chain is reversible with respect to $\mu$ (\cite{haggstrom2002finite}).
So if we can show that $P^\pi_iQ_{i,j} = P^\pi_jQ_{j,i}$, the theorem follows.
This is the standard approach which derives the Metropolis-Hastings algorithm (\citep{metropolis1953equation}).

First, suppose $P^\pi_i|\Omega_j| > P^\pi_j|\Omega_i|$. Then, by inspecting lines 1-5 in the algorithm we see that
\begin{equation*}
\begin{split}
P^\pi_i Q_{i,j} &= P^\pi_i \frac{1}{|\Omega_i|} \frac{\bar{P^\pi_j} |\Omega_i|}{\bar{P^\pi_i} |\Omega_j|} \\
 &= \frac{P^\pi_j}{|\Omega_j|} = P^\pi_j \frac{1}{|\Omega_j|} \\
 &= P^\pi_jQ_{j,i}
\end{split}
\end{equation*}
where the last equation holds because $P^\pi_i|\Omega_j| > P^\pi_j|\Omega_i|$ implies $z = 1$ for the transition from $j$ to $i$. The same applies to the case where $P^\pi_i|\Omega_j| < P^\pi_j|\Omega_i|$.
Finally, suppose $P^\pi_i|\Omega_j| = P^\pi_j|\Omega_i|$. Then,
\begin{equation*}
\begin{split}
P^\pi_i|\Omega_j| &= P^\pi_j|\Omega_i| \\
\implies \frac{P^\pi_i}{|\Omega_i|} &= \frac{P_j^\pi}{|\Omega_j|} \\
\implies P^\pi_j Q_{j,i} &= P^\pi_iQ_{i,j}
\end{split}
\end{equation*}
So the TTCG Gibbs sampler is reversible with respect to $P^\pi$.
\end{proof}